# NANO-ACTUATOR CONCEPTS BASED ON FERROELECTRIC SWITCHING

Ananya Renuka Balakrishna[1], John E. Huber[1], Chad M. Landis[2]

[1]Department of Engineering Science, Parks Road, University of Oxford, Oxford OX1 3PJ
[2]Department of Aerospace Engineering and Engineering Mechanics, University of Texas at Austin, Austin TX 78712-0235

**Abstract**

The concept of a nano-actuator that uses ferroelectric switching to generate enhanced displacements is explored using a phase-field model. The actuator has a ground state in the absence of applied electric field that consists of polarized domains oriented to form a flux closure. When electric field is applied, the polarization reorients through ferroelectric switching and produces strain. The device is mechanically biased by a substrate and returns to the ground state when electric field is removed, giving a repeatable actuation cycle. The mechanical strains which accompany ferroelectric switching are several times greater than the strains attained due to the piezoelectric effect alone. We also demonstrate a second design of actuator in which the displacements are further increased by the bending of a ferroelectric beam. Phase-field modelling is used to track the evolution of domain patterns in the devices during the actuation cycle, and to study the design parameters so as to enhance the achievable actuation strains.

Keywords: ferroelectric, actuator, phase-field

## Introduction

A major drawback of piezoelectric actuators is the limited strain generated by the piezoelectric effect [1],[2]. Consequently it is an attractive idea to exploit the much greater strains that accompany ferroelectric switching for actuation [3],[4]. Two problems prevent this approach from gaining widespread use: Firstly, the electric fields needed for switching would necessitate the use of high voltage in macro-scale devices. Secondly, the strain enhancement comes from 90° ferroelectric switching [3]–[6], but cyclic application of electric fields produces mostly 180° switching. At the micro or nanoscale, the first problem disappears [2], and the second could be resolved through a mechanical interaction with the underlying substrate. For example a state of pre-stress could produce in-plane domains that switch out of plane only when electric field is applied.

In the present work, we study a nano-actuator concept based on ferroelectric switching in $BaTiO_3$. The device concept relies upon thin film technologies [7]–[9] capable of fabricating films of the order of 10nm thickness with comparable feature sizes [10]. At this scale, the ferroelectric element may contain a few domain walls at most [11], and special patterns of domains forming flux closures or vortex-like structures have been predicted [12]–[14]. Research on these nanoscale domain patterns has intensified in recent times due to their promise as ferroelectric devices [15]. The role of factors such as strain [16], [17], electric field [18], and mechanical loads [19] in domain evolution has been investigated, showing that, in principle, nanoscale domains could be controlled so as to produce memory, sensor or transducer type devices [12]–[14], [19]–[23]. There has also been considerable interest in influence of size [11], [24]–[30] on ferroelectric behaviour, consistent with a trend towards miniaturisation of devices. However, there has so far been relatively little progress towards actuation at the nanoscale using ferroelectric switching, which is the objective of the present work.

While a variety of modelling methods, including first principles studies [31], [32], Monte Carlo simulation [30], and Landau-Lifshitz-Gilbert models [25] have been employed to understand ferroelectric properties at the nanoscale, phase-field models [11], [16], [27]–[29], [33]–[36] offer particular advantages when simulating a ferroelectric device. This is because the phase-field approach enables tracking of domain walls, and hence the evolution of domain patterns, in a relatively straightforward and robust way. Here, we simulate domain structure evolution using a previously developed phase-field model [33]. The spirit of the investigation follows work towards engineering or designing effective devices at the nanoscale.

The aim is to demonstrate the working principle of a ferroelectric nano-actuator. Two designs are then explored further to increase the achievable actuation strains. In the first design, named "embedded" actuator, a monocrystalline $BaTiO_3$ thin film is envisaged, which has been deposited within a channel in a pre-stressed substrate. Upon relaxing the substrate a known in-plane strain is imposed upon the ferroelectric film. We select this pre-strain such that the ground state of the ferroelectric forms a polarization flux closure with mainly in-plane domains. This is defined to be the initial state of the actuation cycle. Actuation is then driven by applying an electric field using surface electrodes. The ground state is recovered when electric field is removed. A second design, named "bending" actuator, uses a slender beam-like element of monocrystalline $BaTiO_3$ with multiple electrodes to produce a domain pattern that deforms the actuator into an arched shape. This achieves greater displacements than the embedded actuator, at the cost of reduced actuation force.

**Embedded actuator**

The embedded actuator is simulated using a phase-field model developed by Su and Landis [33] that has been previously applied to study domain wall/defect interaction [33],[34] domain structure energetics [35], effects of strain and temperature in ferroelectric materials [36] and to design nanoscale devices [14], [21]–[23]. This model is based on a generalised Ginzburg-Landau equation:

$$\left(\frac{\partial \psi}{\partial P_{i,j}}\right)_{,j} - \frac{\partial \psi}{\partial P_i} = \beta \dot{P}_i, \qquad \beta \geq 0, \quad (1)$$

where $\psi$ is the Helmholtz free energy, a function of strain ($\varepsilon_{ij}$), electric displacement ($D_i$), the order parameter, which is polarization ($P_i$) and polarization gradient ($P_{i,j}$). In addition to solving equation (1), the equations of mechanical equilibrium, electrical equilibrium and microforce balance are solved as auxiliary conditions, see [33] for further details on the theory, and [34] for the specification of corrected material parameters for $BaTiO_3$. The polarization viscosity $\beta$ controls the rate of evolution of the polarization state. However, in the present work, the evolution of the domain pattern is tracked through equilibrium configurations with $\beta = 0$. Non-zero $\beta$ values are used only in intermediate steps towards equilibrium, or where an equilibrium configuration cannot be found due to instability. The governing differential equations are solved using finite element methods, and typically the element size is chosen such that 180° domain walls span four elements.

The embedded actuator concept is shown in cross-section in Fig. 1. The device consists of a layer of $BaTiO_3$ of length $L$ and thickness (height) $H$, sandwiched between upper and lower electrodes and rigidly constrained by surrounding material on three sides. The electrodes are modelled as thin compliant conductors that do not constrain the actuator. An in-plane pre-strain $\varepsilon_{xx}$ is used to bias the device in favour of in-plane domains when no electric field is applied. The device is modelled in a state of generalized plane strain in the $x - y$ plane such that the strain in the $z$-direction is equal to the lattice a-strain. This corresponds to assuming a depth in the $z$-direction much greater than the cross-sectional dimensions, and is consistent with mechanical clamping of the ferroelectric element by the substrate.

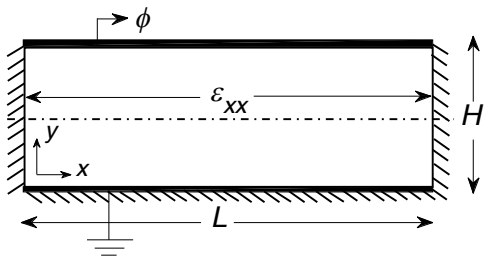

Fig. 1: Schematic representation of boundary conditions on the embedded actuator.

Note that this condition makes polarization in the $z$-direction unfavourable and so enforces polarization in the $x-y$ plane. An actuation cycle starts from an equilibrium ground state with zero voltage on the upper and lower electrodes. The cycle proceeds by increasing voltage $\phi$ on the upper electrode until actuation is achieved and then reducing $\phi$ to zero. Actuation during the cycle is defined by an average displacement of the upper surface of the device. The embedded actuator is representative of $BaTiO_3$ deposited on a lower electrode in a channel etched into the substrate, and capped with an upper electrode. The pre-strain could then be achieved by compressive pre-stress in the substrate during the deposition stage, and subsequent relaxation of the substrate to stretch the device. A similar device could be produced from a continuous thin film on a conductive substrate, with an upper electrode covering a portion of length $L$. For simplicity the fringing of electric field at the device edges and compliance of the surrounding material are neglected.

To demonstrate the actuation cycle, an embedded actuator of size $L = 45$nm and $H = 15$nm with in-plane pre-strain $\varepsilon_{xx} = 0.82\%$ relative to the initial cubic state of the crystal is modelled. Note that the 180° domain wall thickness that arises from the model is approximately 2nm. Electric field $E_y = -\phi/H$ is applied along the negative $y$-axis. When $E_y = 0$, an equilibrium configuration is found with domains oriented to form a flux closure. This is taken to be the ground state of the actuation cycle, see Fig. 2(a). As the magnitude of $E_y$ is increased, the polarization begins to reorient so as to align with the applied field. This switching process is accompanied by ferroelectric strains which displace the top electrode of the embedded actuator.

Fig. 2 shows the domain evolution in the embedded actuator, wherein domains aligned with the electric field are seen to grow at the expense of neighbouring domains Fig. 2(b, c). At $E_y = 50$MV/m a herring-bone-like domain pattern is formed, Fig. 2(d), and on increasing the electric field to $E_y = 55$MV/m a banded domain evolves, Fig. 2(e, f). At a critical value of electric field, $E_y = 85$MV/m, the actuator makes a transition through an unstable needle-like pattern of domains, Fig. 2(g, h), and settles to a uniformly polarized single domain, Fig. 2(i), the actuated state of the device. Note that the field strength required for full switching is much less than the breakdown strength of thin-film barium titanate [37]. An actuation strain $\varepsilon_A$ can be defined, relative to the ground state, as the ratio of average displacement of the top electrode to the device height. Equivalently, the actuation strain $\varepsilon_A$ is the volume average of the strain change from the ground state. In the actuated state, $\varepsilon_A \sim 0.45\%$. Further strengthening of the electric field causes a rise in actuation strain through the piezoelectric effect.

It is of interest to note the stress levels in the actuator. These are typically of order a few hundreds of MPa, comparable to residual stresses in ferroelectric films. However, in the fully actuated state a tensile stress of about 2GPa develops in the $x$-direction. This is due to the substrate constraining the device and preventing contraction in the $x$-$z$ plane. The tensile stresses will act to return the device to the ground state after unloading.

On unloading the electric field from its peak value, the embedded actuator retains the single domain state until $E_y = 5$MV/m. The actuation strain and polarization along the $y$-axis gradually decrease, see Fig. 2(i, j). At $E_y = 5$MV/m, the system goes out of equilibrium and needle like domains with in-plane polarization nucleate, Fig. 2(j). This evolves into a herring-bone-like domain pattern and finally returns the actuator to its initial vortex-like flux closure state, completing the actuation cycle, Fig. 2(k, l). Note that there were three transitions through unstable states during the actuation cycle, at $E_y = 55$MV/m, $E_y = 85$MV/m during loading and at $E_y = 5$MV/m during unloading. In each case the finite element solver did not find equilibrium solutions and the transition was modelled by using non-zero $\beta$ values to relax the system towards equilibrium. However, we found that the exact route taken during such unstable transitions was highly sensitive to the choice of $\beta$ values, and this could affect the overall pattern of domains. Effectively the $\beta$ values used control the rate at which the system passes through an instability. Furthermore the system could also settle into an anticlockwise vortex state symmetrically opposite to the clockwise vortex of Fig. 2(l). This sensitivity in the exact route followed by the pattern of domains did not significantly affect the actuation strain or the ability of the device to operate cyclically.

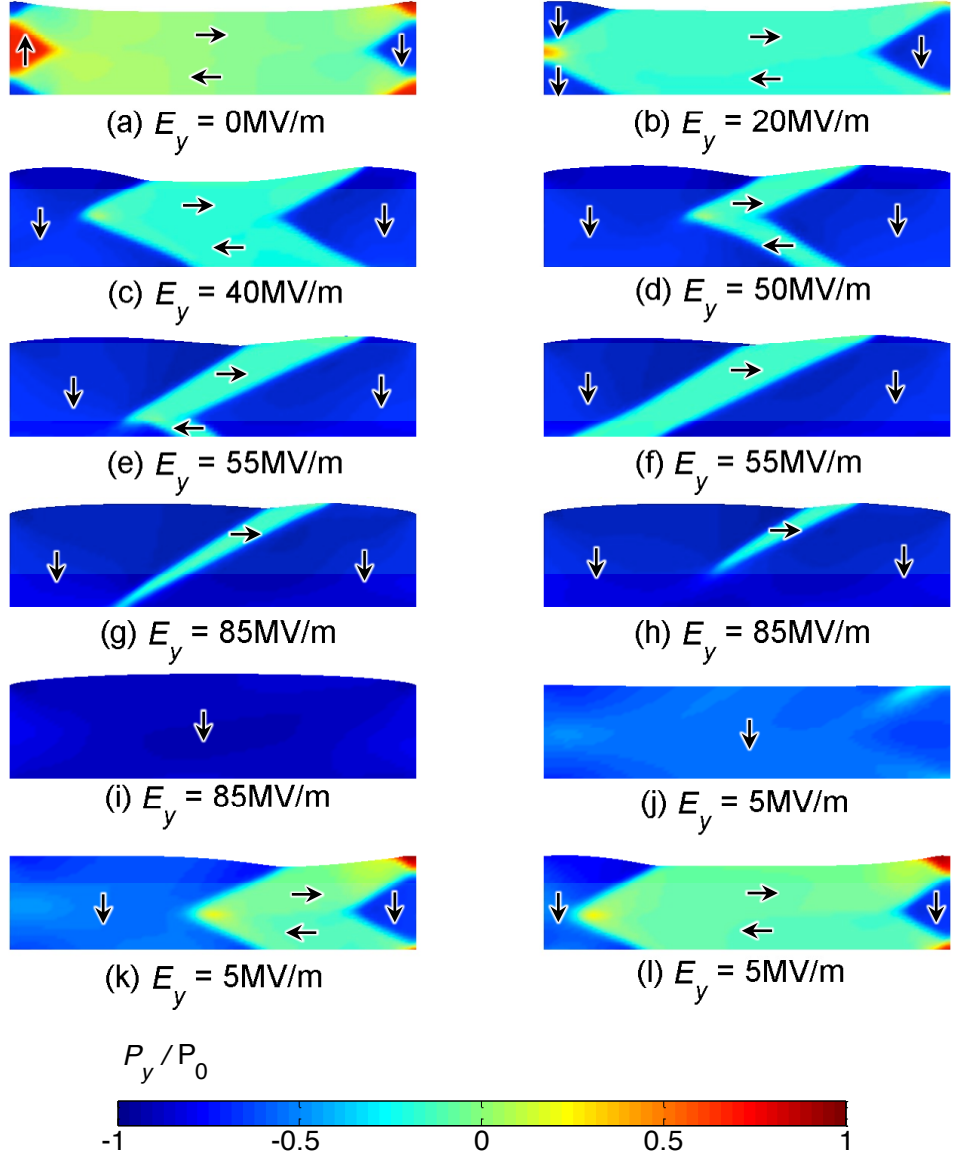

Fig. 2: Actuation cycle in the embedded actuator, showing polarization along the $y$-direction. $P_0 = 0.26 C/m^2$. (The deformed configuration is shown with displacements exaggerated.)

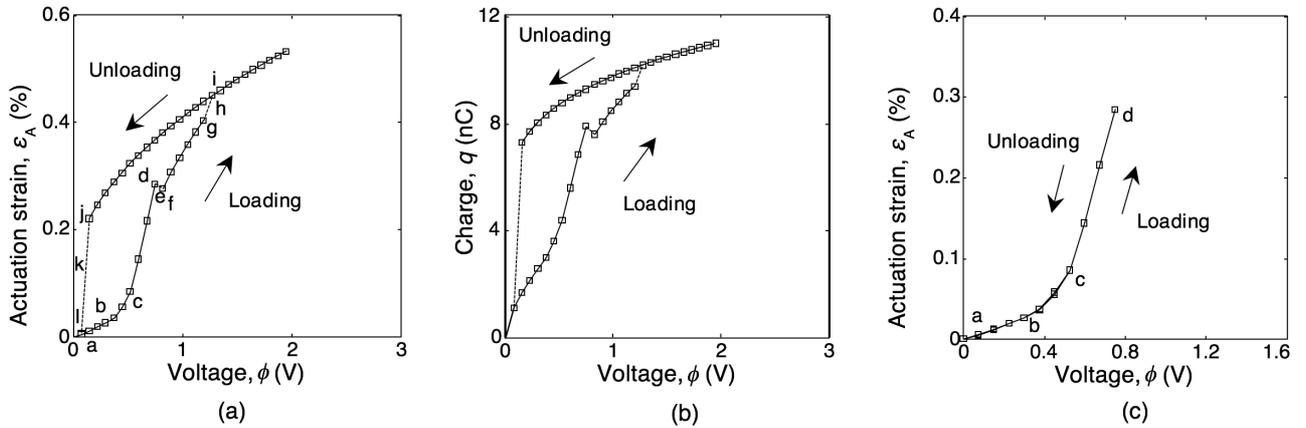

Fig. 3: Embedded actuator (a) Voltage-strain curve with complete domain switching (b) Corresponding voltage-charge curve (c) Voltage-strain curve with partial switching. Labels correspond to the states shown in Fig. 2.

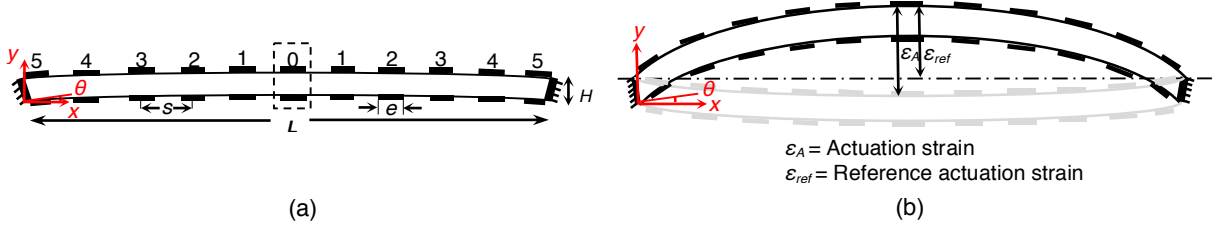

Fig. 4: (a) Schematic diagram of bending actuator. The dotted boundary indicates an electrode pair and the electrode pairs are numbered $k = 0 ... 5$. (b) Representation of strains defined for the bending actuator.

The voltage-strain curve for the embedded actuator is shown in Fig. 3(a), where the unstable transitions are indicated by dashed lines. The total charge on the upper electrode during the actuation cycle is shown in Fig. 3(b). It should be noted that the hysteresis in the voltage-strain and voltage-charge curves arises entirely from the unstable transitions: at all other points, the model is incrementally moving between equilibrium states that are "nearby" in the sense that the changes in domain pattern are also incremental. Consequently, by choosing a peak value of applied electric field less than 55MV/m, the device can be operated without hysteresis. This is illustrated in Fig. 3(c), where the applied electric field was limited to $E_y = 50$MV/m thereby avoiding unstable transitions. The loading curve in Fig. 3(c) is identical to the corresponding part of Fig. 3(a), while the unloading curve follows the loading curve. The voltage-strain curve then shows no hysteresis to within the model precision, and results in a peak actuation strain of 0.28%.

### Bending actuator

The actuation displacements can be enhanced by exploiting the large sideways displacement attained on bending a slender beam fixed at its ends. A sideways displacement is achieved by inducing a positive strain along the beam length. Based on this concept, a design for a bending actuator of nominal dimensions 365nm length × 15nm height is shown in Fig. 4(a). The ends of the actuator are rotated by an angle $\theta = 3°$ and clamped in this state, inducing an initial curvature that stabilises bending towards the $+y$-direction. These mechanically clamped edges are maintained at 5V throughout the actuation cycle. Electrodes of width $e = 17$nm are modelled in pairs along the top and bottom surfaces of the actuator with a nominal gap of $s = 35$nm measured between the centres of adjacent electrodes. The actuation strain, $\varepsilon_A$ for the bending actuator is defined as ratio of the average displacement of the $k = 0$ electrode to the device height of 15nm. For future discussion we also introduce a reference actuation strain, $\varepsilon_{ref}$ which is defined as the ratio of net displacement of the $k = 0$ electrode from a reference line located at $y = 15$nm, to the device height, see Fig. 4(b).

The electric potential applied to the upper electrodes ($\phi_U$) and lower electrodes ($\phi_L$) are of the following form:

$$\phi_U = \phi_0 + k\phi_1 \qquad (2)$$

$$\phi_L = \alpha(\phi_0 + k\phi_1) \qquad (3)$$

where $k$ is the electrode pair number shown in Fig. 4 and voltages $\phi_0$, $\phi_1$ are constants with values of 2.5V and 0.5V respectively. The loading parameter, $\alpha$, is varied from 0 to 1, taking the actuator between the "off" and "on" states respectively. By controlling $\alpha$, a nominal electric field $E_{nom} = \alpha\phi_1/s$ is set up along the length of the actuator. For later use the voltage $\phi_{nom} = \alpha\phi_1$ is introduced, representing the net potential difference between adjacent electrodes on the lower surface. The effect of this distribution of electrode voltages is to produce an average electric field in the $-y$-direction in the "off" state, and an electric field in the $\pm x$-direction in the "on" state; this activates 90° switching.

When $E_{nom} = 0$, the bending actuator settles into a ground state with polarization pattern as shown in Fig. 5(a). On increasing $E_{nom}$, domains aligned with the $\pm x$-direction nucleate and grow, see Fig. 5(b-d). Ferroelectric switching extends the actuator along its length, but the presence of end constraints forces this extension to be accommodated by bending. At a critical value of electric field $E_{nom} = 14.3$MV/m the simulation makes an unstable transition to the fully actuated state, Fig. 5(e-f), resulting in actuation strain $\varepsilon_A \sim 15\%$.

On reducing the value of applied electric field to $E_{nom} = 11$MV/m there is a rapid decrease in the actuation strain, Fig. 5(g-h). New domains nucleate from the device centre and spread towards the clamped edges, Fig. 5(i, j). The appearance of 90° domains at the ends of the beam in Fig. 5(g, h) begins to flatten the device returning it to its ground state, completing the actuation cycle. Note that the bending actuator can settle into a ground state with flux closures oriented in opposite sense, Fig. 5(j), in comparison to the initial state of the actuator, Fig. 5(a). Thus the actuation cycle need not be repeatable in terms of the domain pattern formed, however this does not significantly affect the actuation strain. A voltage-strain curve for the bending actuator is shown in Fig. 6(a). In order to show the voltage-charge behaviour of this actuator, we define a work-equivalent charge $q_{\text{nom}}$ such that $\phi_{\text{nom}} \delta q_{\text{nom}} = \int_S \phi \delta q \, ds$, where $\phi$ is the surface voltage, and $\delta q$ is an increment of surface charge density, with the integral conducted over the area of the surface electrodes. Fig 6(b) shows the work-equivalent charge during the actuation cycle. This shows a steep increase in charge early in the loading cycle ($\phi_{\text{nom}} < 0.1V$), indicating that the polarization state is highly mobile. This is consistent with the presence of non-zero electrode voltages and hence the presence of electric fields even in the "off" state. Small changes in voltage result in domain wall movement. Fig. 6(b) also shows hysteresis in the voltage-charge cycle indicating energy dissipation during the unstable transitions.

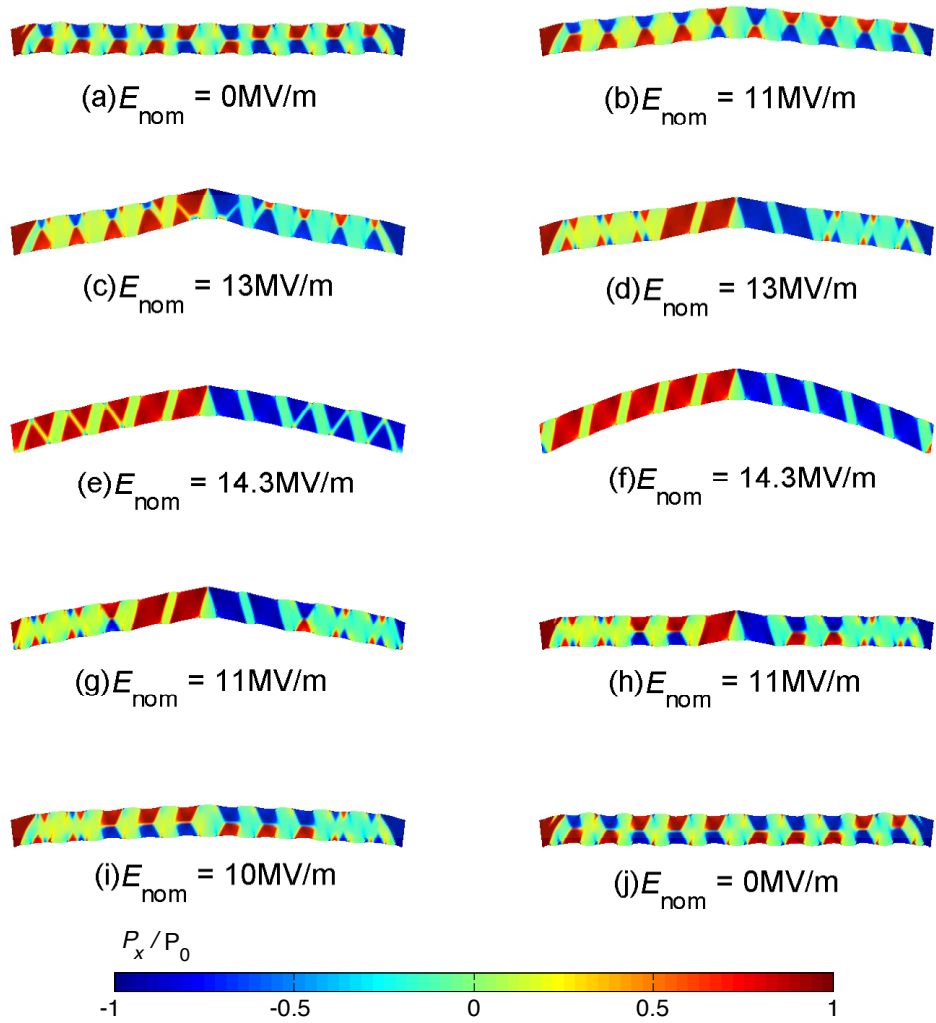

Fig. 5: Actuation cycle in the bending actuator, showing polarization in the *x*-direction.

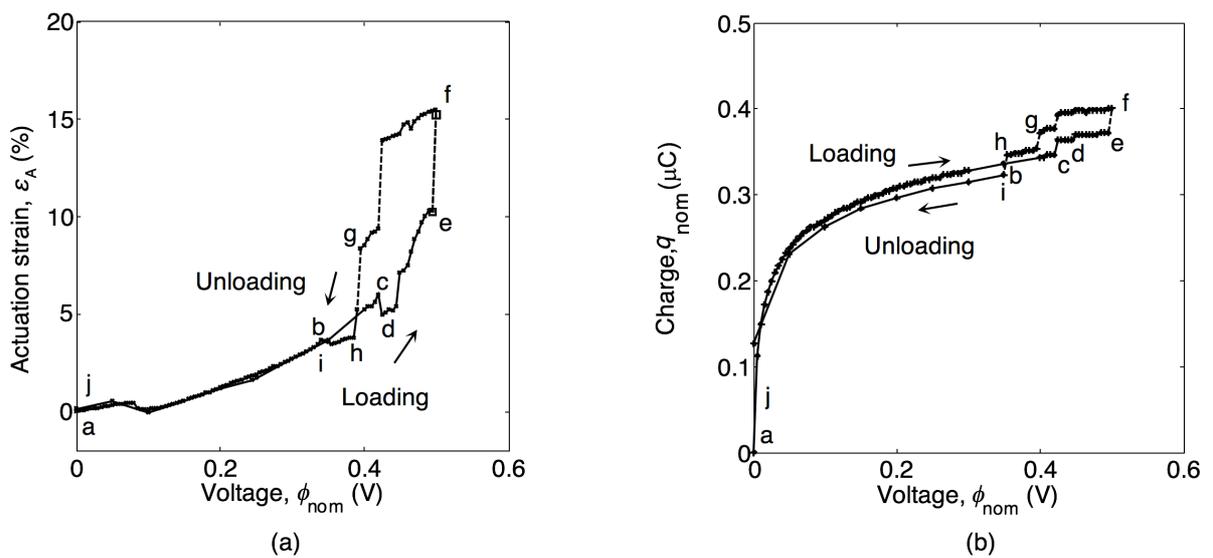

Fig. 6: Bending actuator (a) Voltage-strain curve (b) Voltage-charge curve. Labels correspond to the states shown in Fig. 5.

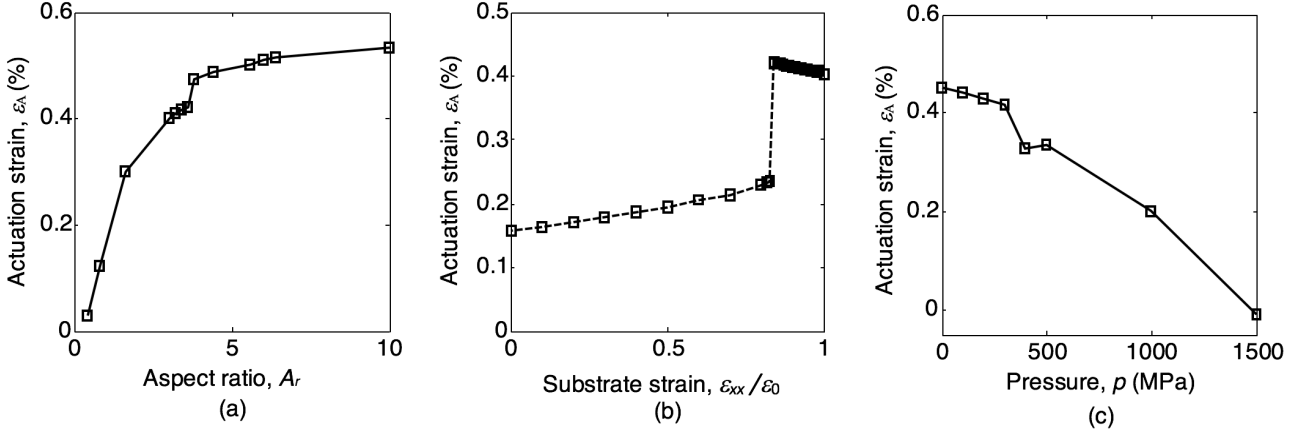

Fig. 7: Parametric study for the embedded actuator showing the effect of varying (a) Aspect ratio, $A_r$ (b) In-plane substrate strain, $\varepsilon_{xx}$ (c) Uniform pressure, $p$.

**Parametric Study**

The actuators can be improved by exploring the space of design parameters to enhance the achievable actuation strains. In the case of the embedded actuator, variation of the aspect ratio $A_r = L/H$ and the imposed in-plane strain $\varepsilon_{xx}$ are considered. Both could be controlled through the manufacturing process. For the bending actuator, design parameters include the electrode coverage $e_c = e/s$ and angle of end rotation $\theta$ at the beam ends. Other parameters, including the size or scale of the devices and the distribution of voltages on the electrodes should be considered for a full optimization. However, we found that increasing the device size by a factor of 3 or more typically resulted in the formation of more complex domain patterns that were not necessarily repeatable upon cycling.

First consider varying the aspect ratio $A_r$ of the embedded actuator, with height held constant at $H = 15$nm. Fig. 7(a) plots the resulting actuation strains $\varepsilon_A$ attained on ferroelectric switching versus aspect ratio. Aspect ratios in the range 0.4 to 1 result in low values of actuation strain, typically $\varepsilon_A < 0.1\%$. These device designs are effectively clamped against actuation by the surrounding substrate material. On increasing the aspect ratio to $A_r = 3$ a steep rise in actuation strain to $\varepsilon_A = 0.45\%$ is observed: in this regime the device functions as illustrated in Fig. 2, and increasing $A_r$ steadily reduces the influence of end constraint. Beyond $A_r \sim 3.5$, further increase in $A_r$ does not contribute a significant increase of actuation strain, which saturates at about $\varepsilon_A = 0.5\%$. In this regime, the actuator did not reliably return to the ground state upon unloading. The small step increase in actuation strain when $A_r = 3.5$ results from an increase in the applied electric field required to reach the fully actuated state. This was caused by a change in the domain patterns that form during the transition to the single domain state. To summarise, it was found that $A_r \sim 3$ provides a satisfactory compromise, giving a high value of actuation strain, while maintaining a stable actuation cycle.

It is interesting to note that at aspect ratios $A_r \sim 0.4$ ($L = 6$nm, $H = 15$nm) the simulated device did not return to the ground state when the upper electrode voltage $\phi$ was reduced to zero. Instead, a single domain state (polarized in the $-y$ direction) was retaining after unloading. A negative voltage pulse could then return the device to the ground state. However a stronger negative voltage pulse could flip the polarization into a stable single domain state in the $+y$ direction. This design thus has potential as a tri-state memory element, though it should be noted that the device length was here only a few domain wall widths, which could present practical difficulty for manufacture.

Next consider varying the in-plane residual strain $\varepsilon_{xx}$, while keeping $A_r = 3$. The strain $\varepsilon_0 = 0.0082$ corresponds to the spontaneous strain of BaTiO$_3$ in the tetragonal phase. We set this value as a maximum residual in-plane strain value and vary $\varepsilon_{xx}/\varepsilon_0$ in the range zero to unity. When $0.84 < \varepsilon_{xx}/\varepsilon_0 < 1$ the embedded actuator behaves as indicated in Fig. 2. There is a maximum actuation strain of 0.46% occurring when $\varepsilon_{xx}/\varepsilon_0 = 0.84$, but there is relatively little gain from varying $\varepsilon_{xx}$ within this range, see Fig. 7(b). On reducing the in-plane residual strain below $0.84\varepsilon_0$, the actuator remains in a uniformly polarized single domain state during cycling and so the strain enhancement due to ferroelectric switching is lost. This indicates the importance of the substrate strain in stabilising in-plane domains in the ground state.

Finally, the response of the embedded actuator with $A_r = 3$ and $\varepsilon_{xx}/\varepsilon_0 = 1$ to uniform pressure $p$ on its upper surface is studied. Fig. 7(c) shows the actuation strains achieved for a range of pressure $p = 0 - 1500$MPa. In practice applied pressure > 500MPa may be difficult to achieve. However, the simulations allow us to find a theoretical blocking pressure, $p \sim 1500$MPa, at which $\varepsilon_A$ drops to zero. Since the actuation strain is not greatly affected by applied pressure in the range 0-500MPa, the embedded actuator can be effective in applications where working against a force or pressure is required.

We now carry out a parametric study of the bending actuator, by exploring the effect of variations in the electrode coverage $e_c$, and the angle of end rotation $\theta$.

To study the effect of electrode coverage, $e_c$ was varied in the range 0.2 - 0.8. Values beyond this range would require features (electrodes or gaps in electrodes) smaller than 5nm which we take as a practical limit for manufacture. To explore this range, the "on" and "off" states of the device were separately traced through a virtual process of slow electrode growth on the actuator surface. The pattern of voltages $\phi_U$ and $\phi_L$ was held with either $\alpha = 0$ or 1 while increments in $e_c$ were produced by increasing the electrode width, $e$. This allowed the actuation strain for each design to be estimated without simulating the entire actuation cycle. These estimates are open to significant error because the complex domain patterns present in the actuated state are not necessarily reproduced if actuation is achieved by a different route. For a check on the reliability of the estimates, the full actuation cycle from ground state to actuated state was simulated at selected $e_c$ values. While small differences in domain pattern of the actuated state were observed, the estimated actuation strains were generally within ±3% of the full-cycle value.

To understand the variation in behaviour, note that in the "off" state, domains polarized parallel to the y-axis form between the upper and lower electrode of each pair. These domains tend to shorten the beam length giving a flat initial state. But when the electrode coverage is low ($e_c = 0.2$ to 0.35), a combination of triangular and banded domains appear in the ground state. This domain pattern provides an initial upward curvature to the "off" state of the device and the actuation strain, which measures the change between "off" and "on" is small. Increasing the electrode coverage increases the actuation strain, up to about $e_c = 0.5$, at which point the actuator cycles between a nearly flat "off" state and a curved "on" state. Further increase in $e_c$ does not significantly affect the actuation strain. Fig. 8(a) shows the resulting actuation strain versus electrode coverage.

We next explore variation of the angle of end rotation $\theta$, holding $e_c$ constant at 0.45. Once again, the "on" and "off" states were traced by holding $\phi_U$ and $\phi_L$ constant, this time making incremental changes to the positions of nodes at the beam ends to simulate rotation of the end supports, changing $\theta$ in the range 0°- 6°, the upper limit corresponding to the beam adopting an arc shape with the upper surface strained by $\varepsilon_0$ relative to the lower surface. Varying $\theta$ did not have a significant effect on the actuation strain of the device. However, the reference actuation strain, $\varepsilon_{\text{ref}}$ which accounts for net displacement of the bending actuator from a reference line at $y = 15$nm linearly increases with $\theta$, see Fig. 8(b). For $\theta < 3°$ the actuator adopts a downward curvature in the ground state while at $\theta = 3°$ the beam is nearly flat; it is curved upward for $\theta > 3°$. Hence the values of $\varepsilon_{\text{ref}}$ and $\varepsilon_A$ are approximately the same when $\theta = 3°$ and this configuration is chosen for further study.

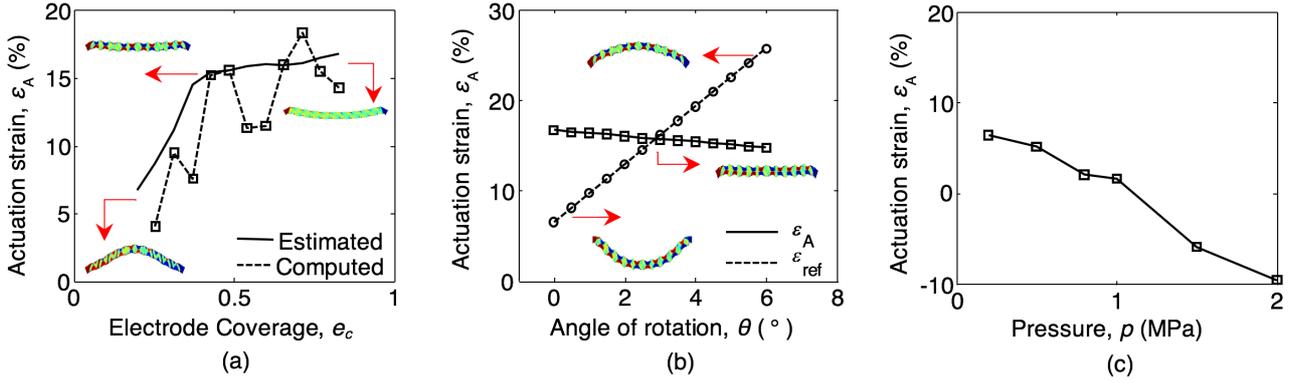

Fig. 8: Selection of design parameters for the bending actuator: (a) Variation of electrode coverage, $e_c$ (b) Variation of angle of end rotation, $\theta$ (c) Actuation strain when working against pressure $p$. Inset schematics show the ground state of the actuator.

Finally, the response of the actuator working against a uniform pressure $p$, applied over the upper surface, was tested. The slender design of the bending actuator limits its capability to work against loads: at $p = 1$MPa, the achievable actuation strain is reduced to $\varepsilon_A \sim 1\%$ and further increase in the load blocks the actuator, see Fig. 8(c). This confirms that the bending actuator is best suited to applications that require displacement at negligible loads.

## Conclusion

Actuation based on ferroelectric switching has been demonstrated through two designs of nano-actuators using a phase-field model as a design tool. The domain evolution during the actuation cycle was visualized by plotting the polarization patterns in the actuators. The embedded actuator, which was driven by electric field and reversed by mechanical clamping, resulted in actuation strains of 0.45% and worked against pressure loads of up to at least 500MPa. Greater displacements, of 15% the device height, were achieved in the bending actuator. Design parameters such as aspect ratio, in-plane strain $\varepsilon_{xx}$ were studied for the embedded actuator to enhance the achievable displacements, while the effect of electrode coverage $e_c$ and angle of end rotation $\theta$ were explored for the bending actuator. The mechanism of actuation presented indicates that actuation strains resulting from ferroelectric switching are several times larger than the piezoelectric strains alone and could potentially be exploited in nanoscale devices.

## Acknowledgement

The authors wish to thank Dr D. Carka for advice on modelling. This work was an outcome from EPSRC grant EP/G065233/1 and NSF grant DMR-0909139. Ananya Renuka Balakrishna acknowledges the support of a scholarship from the Felix Trust. Chad M. Landis also acknowledges support from NSF grant CMMI-1068024.